\documentclass[12pt]{article}
\usepackage[dvips]{graphicx}
\usepackage{amssymb, amsmath, mathbbol}
\usepackage[centerlast,footnotesize]{caption2}

\newcommand{\lapprox}{\raisebox{-0.5ex}{$\
\stackrel{\textstyle<}{\textstyle\sim}\ $}}
\newcommand{\gapprox}{\raisebox{-0.5ex}{$\
\stackrel{\textstyle>}{\textstyle\sim}\ $}}

\usepackage{graphicx}
\newcommand{\One}{1\kern-4.5pt1}

\newcommand{\be}{\begin{equation}}
\newcommand{\ee}{\end{equation}}

\def\lesim{${\lower 2pt\hbox{$\scriptstyle
<$}\atop\raise 4pt\hbox{$\scriptstyle\sim$}}$} 
\def\grsim{${\lower2pt\hbox{$\scriptstyle >$} \atop\raise4pt\hbox 
{$\scriptstyle\sim$}}$} 

\begin{document}
\begin{center}
\begin{flushright}
September 2015
\end{flushright}
\vskip 10mm
{\LARGE
Strong Interaction Effects at a Fermi Surface in\\
a Model for Voltage-Biased Bilayer Graphene\\ 
}
\vskip 0.3 cm
{\bf Wes Armour$^{a}$, Simon Hands$^b$ and Costas  Strouthos$^c$}
\vskip 0.3 cm
$^a${\em 
Oxford e-Research Centre, University of Oxford, 7 Keble Road, Oxford OX1 3QG, United Kingdom.}
\vskip 0.3 cm
$^b${\em Department of Physics, College of Science, Swansea University,\\
Singleton Park, Swansea SA2 8PP, United Kingdom.}
\vskip 0.3 cm
$^c${\em  School of Sciences, Department of Computer Science,\\
European University Cyprus, 1516 Cyprus.}
\vskip 0.3 cm
\end{center}

\noindent
{\bf Abstract:} 
Monte Carlo simulation of a 2+1 dimensional model of voltage-biased bilayer
graphene, consisting of 
relativistic fermions with 
chemical potential $\mu$ coupled to charged excitations with opposite
sign on each layer, 
has exposed non-canonical scaling of bulk observables 
near a quantum critical point found at strong coupling.
We present a calculation of the quasiparticle dispersion relation $E(k)$ as a
function of exciton source $j$ in the
same system, employing partially twisted boundary conditions to boost the number
of available momentum modes. The Fermi momentum $k_F$ and superfluid gap
$\Delta$ are extracted in the $j\to0$ limit for three different values of $\mu$, and
support a strongly-interacting scenario at the Fermi surface with $\Delta\sim
O(\mu)$.  We propose an explanation for the observation $\mu<k_F$ in terms of a
dynamical critical exponent $z<1$.

                                                                                
\vspace{0.5cm}

\newpage
\section{Introduction}

There are very few many-body systems 
permitting Monte Carlo simulation without
the need to confront a Sign Problem. In Ref.~\cite{Armour:2013yk} we introduced a
new member to this class, based on an effective theory of bilayer graphene in
which charge carrying excitations are modelled as $N_f=4$ relativistic fermions moving
in a 2$d$ plane. The introduction of chemical potential $\mu$ is via a bias voltage in the perpendicular
direction which induces equal densities of electrons on one layer and holes on the
other; this is analogous to isospin chemical potential in QCD and yields a
real positive fermion determinant amenable to orthodox Monte Carlo methods. 
The simulation was performed in the vicinity of a quantum critical point (QCP)
found at strong coupling, and the main result was the demonstration that the
ground state is a superfluid formed by condensation of electron-hole exciton
pairs, and that the response to chemical potential is governed by the
non-canonical scaling forms (\ref{eq:nscaling}) and (\ref{eq:condscaling})
given in Sec.~\ref{sec:form}.

The model~\cite{Armour:2013yk} as originally devised for bilayer graphene~\cite{Castro} is artificial in a few
respects. Firstly, the description in terms of $N_f=4$ relativistic species (ie.
$N_f=2$ electrons and $N_f=2$ holes) is
only justified by the band structure of the tight-binding model 
in the presence of an inter-layer ``skew'' coupling 
breaking the trigonal symmetry of the underlying lattice~\cite{strain}. Secondly, the interaction between
charge densities is simplified to be a local four-fermi contact, although a
more realistic unscreened Coulomb interaction can be modelled with the
introduction of a third spatial lattice direction to capture the 
electrodynamics~\cite{D&L}. Finally, intra- and inter-layer interactions have the same
coupling strength, as a necessary condition of keeping the fermion determinant
real. Nonetheless, it shares the essential features of a more general
model for double-layer graphene systems in which there is some
hybridization permitting interlayer tunnelling~\cite{SPM}. With $N_f=1$ this
approach is also
applicable to surface states of topological insulators, motivating study with
variable $N_f$~\cite{Hands:2008id}.

The results of \cite{Armour:2013yk} were interpreted in terms of strong
interaction effects at a Fermi surface; if exciton pairs within a shell of
thickness $\Delta$ condense around the Fermi surface centred at $k_F$, then the
anomalous scaling (\ref{eq:nscaling},\ref{eq:condscaling})
is consistent with a BCS mechanism with $\Delta\sim O(\mu)$. Everything is
to be viewed in the context of an effective field theory valid near the QCP. In order to
put this picture on firmer footing, and also to expose the Fermi surface, in
this paper we use Monte Carlo simulation to calculate the quasiparticle dispersion relation $E(k)$,
identifying $k_F$ with the location of the minumum and $\Delta$ with $E(k_F)$.
The main results are summarised in Fig.~\ref{fig:gap} below.
In Sec.~\ref{sec:form} we present the model and review the main
findings of \cite{Armour:2013yk}, then in Sec.~\ref{sec:quasi} present the
calculation of $E(k)$. Our results, summarised in Sec.~\ref{sec:discussion},
indeed support the picture of a Fermi surface disrupted by strong interactions,
leading to the formation of a gap $\Delta$ increasing monotonically with $\mu$.
We also discuss our observation of the striking inequality $\mu<k_F$, 
and propose an explanation in terms of an estimate for the dynamical critical exponent
$z<1$.

\section{Formulation and Simulation of the Model}
\label{sec:form}

The model we use to describe voltage-biased bilayer graphene in terms of $N_f=4$
relativistic fermions is described by the following
Lagrangian~\cite{Armour:2013yk}:
\begin{equation}
{\cal L}=(\bar\psi,\bar\phi)\left(\begin{matrix}D[V;\mu]&ij\cr
-ij&D[V;-\mu]\cr\end{matrix}\right)\left(\begin{matrix}\psi\cr\phi\end{matrix}\right)+{1\over{2g^2}}V^2
\equiv\bar\Psi{\cal M}\Psi+{1\over{2g^2}}V^2.
\label{eq:Sbilayer}
\end{equation}
Here $\psi$ and $\phi$ are 4$\times$2-component spinors each describing two
Dirac flavors, with $\psi,\bar\psi$ corresponding to electron degrees of
freedom on one layer and $\phi,\bar\phi$ holes on the other; 
$V$ is an auxiliary field defined on the timelike links of
the lattice which approximates an ``instantaneous'' Coulomb potential
governed by 3+1$d$ Maxwell electrodynamics; and $j$ a symmetry-breaking gap
parameter due to interlayer pairing. Because some weak interlayer hybridization
is likely to be present in double-layer systems, in general
$j\not=0$~\cite{SPM};
however we will attempt to 
extrapolate $j\to0$ so that exciton condensation can be viewed as a spontaneous symmetry
breaking U(4)$\otimes$U(4)$\to$U(4)~\cite{Armour:2013yk}. The bias voltage is given by $2\mu$
where $\mu$ is formally equivalent to the isospin chemical potential in QCD. 
In continuum notation the covariant derivative operator is
\begin{equation}
D[V;\mu]=\delta^{\alpha,\beta}\left(
\sum_{\nu=0,\ldots,2}\gamma_\nu\partial_\nu+(iV+\mu)\gamma_0\right)=-D^\dagger[V;-\mu],
\end{equation}
where $\alpha,\beta$ run over $N_f=2$ Dirac flavors. The minimal coupling to $V$
implies that $\psi\psi$, $\phi\phi$ and $\phi\psi$ interactions are all of equal
strength, which is required for the action to be real following integration
over the fermions. This corresponds to the interlayer separation $d\to0$ in the 
double-layer model~\cite{SPM}.

In terms of staggered fermion fields living on the sites $x,y$ of a
2+1$d$ cubic lattice $D$ is written
\begin{equation}
D^{\rm latt}_{x,y}={1\over2}\left[\eta_{0x}e^\mu(1+iV_x)\delta_{y,x+\hat0}
-\eta_{0x}e^{-\mu}(1-iV_x)\delta_{y,x-\hat0}+
\sum_{\nu=1,2}
\eta_{\nu x}
(\delta_{y,x+\hat\nu}-
\delta_{y,x-\hat\nu})\right]
\label{eq:Dlatt}
\end{equation}
where the sign factors $\eta_{\nu x}\equiv(-1)^{x_0+\cdots x_{\nu-1}}$ ensure a
covariant weak-coupling continuum limit. Eq.~(\ref{eq:Dlatt}) was also used to
model electron excitations in monolayer graphene~\cite{Armour:2009vj}.
Note that in 2+1$d$ a single staggered
fermion automatically describes $N_f=2$ continuum flavors~\cite{BB}.  
We work in units in
which both spatial $a_s$ and temporal $a_t$ lattice spacings are set to unity
(equivalent to setting the bare Fermi velocity $v_F=1$);
note that for a non-covariant action there is no
reason {\it a priori} to assume $a_s=a_t$, though since the value of the
ratio $a_t/a_s$ is driven by UV physics it does not depend on $\mu$ and only weakly on
$j$, so may be assumed constant throughout this paper.
On the assumption that the dimension of $D^{\rm latt}$ is even, it is straightforward
to show 
\begin{equation}
\mbox{\rm det}{\cal M}=\mbox{\rm det}[D^\dagger D+j^2]>0
\end{equation}
and hence there is no obstruction to Monte Carlo simulation using orthodox
numerical techniques.

In \cite{Armour:2013yk} we presented results from numerical simulation of the
model (\ref{eq:Sbilayer},\ref{eq:Dlatt}) using a hybrid Monte Carlo algorithm.
The coupling $g^{-2}a^\prime=0.4$, where the factor $a^\prime=a_s^2a_t^{-1}$
follows because the interaction couples charge densities, was chosen in the vicinity of
the quantum critical point (QCP), although since $N_f=4\lapprox
N_{fc}=4.8(2)$~\cite{Hands:2008id} it is
hard to ascertain with confidence on which side of the phase boundary at $\mu=0$
we are sitting. Two principal observables were monitored as a function of $\mu$
and $j$, namely the carrier density
\begin{equation}
n_c\equiv{1\over2}{{\partial\ln{\cal Z}}\over{\partial\mu}}={1\over2}\langle\bar\psi D_0\psi-
\bar\phi D_0\phi\rangle,
\end{equation}
and the exciton condensate
\begin{equation}
\langle\Psi\Psi\rangle\equiv{{\partial\ln{\cal Z}}\over{\partial j}}=
i\langle\bar\psi\phi-\bar\phi\psi\rangle.
\end{equation}
Following extrapolation to the limit $j\to0$ 
(so long as $\mu a_t<0.3$ at which
point saturation artifacts set in and the continuum approximation 
fails),
both observables showed behaviour consistent with rising smoothly from zero 
as $\mu$ is increased, with 
\begin{eqnarray}
n_c&\propto&\mu^{3.32(1)}\label{eq:nscaling}\\
\langle\Psi\Psi\rangle&\propto&\mu^{2.39(2)}\label{eq:condscaling}. 
\end{eqnarray}
This is to be contrasted
with the expectations from weak-coupling. In free field theory the carrier
density depends on the volume contained within the Fermi surface; for
relativistic fermions $k_F\approx\mu$ and hence in 2+1$d$ $n_c\propto\mu^2$.
Similarly, in weak coupling we expect the exciton condensate to arise from
electron-hole pairing with equal and opposite momenta from within a shell of
thickness $2\Delta$ centred on $k_F$; hence
$\langle\Psi\Psi\rangle\propto\Delta\mu$. The non-canonical scaling is
taken to be a symptom of strong field fluctuations near the QCP. Moreover, since
according to Luttinger's theorem $n_c\propto k_F^2$ even in the presence of
interactions, we can adapt this argument to estimate the scaling of the gap
$\Delta(\mu)$:
\begin{equation}
\Delta(\mu)={{\langle\Psi\Psi(\mu)\rangle}\over{n_c^{1\over2}(\mu)}}\propto
\begin{cases}\mbox{constant}&\mbox{weak coupling};\cr
\mu^{1.44(1)}&\mbox{near QCP}.
\end{cases}\label{eq:Deltahyp}\end{equation}
While the numerical value for the exponent should probably not be taken
too seriously, the qualitative difference between the scaling predicted in weak
coupling and observed near the QCP is striking;
indeed, Fig.~13 of \cite{Armour:2013yk} shows the ratio
$\langle\Psi\Psi\rangle/n_c^{1\over2}$ extrapolated to $j\to0$ almost linearly
proportional to $\mu$. Since near a QCP  $\mu$ is the only energy scale in
the problem, naively $\Delta\propto\mu$ is expected. As both arguments contain
assumptions, it is clear that a direct calculation of $\Delta$ from the
quasiparticle propagator $\langle\Psi(0)\bar\Psi(x)\rangle$ is needed.

\section{Quasiparticle Dispersion}
\label{sec:quasi}

While the results outlined in the previous section are intriguing the
conclusions, drawn from simulations of systems with both significant UV and IR
artifacts on a restricted range of $\mu$ values, are necessarily provisional.
In this paper we will present complementary information
through analysis of the quasiparticle dispersion
relation $E(\vec k)$, which will enable identification of the Fermi momentum $k_F$
and direct estimation of the superfluid gap $\Delta$. The basic observables are the
timeslice correlators in momentum space:
\begin{equation}
C_N(\vec k,t)=\sum_{\vec x}\langle\psi(\vec0,0)\bar\psi(\vec x,t)\rangle
e^{-i\vec k.\vec x};\;\;\;
C_A(\vec k,t)=\sum_{\vec x}\langle\psi(\vec0,0)\bar\phi(\vec x,t)\rangle
e^{-i\vec k.\vec x},
\label{eq:props}
\end{equation}
where we distinguish between {\it normal\/} propagation of an electron or hole
within a layer, and {\it anomalous\/} propagation in which an electron moving in
one layer is absorbed by an exciton, transferring its momentum to an electron
moving in the other. On a finite volume in the absence of explicit symmetry
breaking $j=0$, the anomalous
component $C_A$ necessarily vanishes.

Many momentum modes must be available in order to obtain a good resolution for $E(\vec k)$.
Naively this would entail making at least one of the spatial dimensions of the
lattice as large as possible~\cite{Hands:2004uv}. It is much more efficient,
however, to employ partially twisted boundary conditions~\cite{Flynn:2005in} in
which the constraint $\psi(L_x)=e^{i\theta_x}\psi(0)$, with the angle $\theta_x$
adjustable, is implemented in the
calculation of the propagator (\ref{eq:props}) so that accessible modes have
\begin{equation}
k_x={{2\pi n+\theta_x}\over L_x}.
\end{equation}
In practice the twist is implemented via a field redefinition so that each
$x$-link in (\ref{eq:Dlatt}) is multiplied by a phase $e^{\pm i\theta_x/L_x}$.
Treating $+x$ and $-x$ independently, the choice $\theta_x=2\pi/3$ permits an
effective tripling of the number of available modes for a given $L_x$ for the
cost of an extra inversion (although equivalent statistics requires twice as
many twisted inversions as non-twisted). For the two volumes investigated here,
the accessible modes were
$ka_s=0,{\pi\over{48}},\ldots,{{n\pi}\over{48}}\ldots$
($32^3$) and 
$ka_s=0,{\pi\over{72}},\ldots,{{n\pi}\over{72}}\ldots$
($48^3$). In all cases the maximum accessible momentum for staggered fermions is
${\pi\over2}$. 

Results were generated at $g^{-2}a^\prime=0.4$ at $\mu a_t=0.1, 0.2$ on $32^3$
and $\mu a_t
=0.1, 0.15,0.2$ on $48^3$, with exciton source $j\tilde a=0.005,0.01,0.02,\ldots,0.05$.
(here $\tilde a=a_s^\alpha a_t^{1-\alpha}$ with $\alpha$ defined by 
the renormalisation prescription adopted for the non-conserved density
$\bar\psi\phi$). A hybrid Monte Carlo algorithm with $\delta\tau=0.0025$ and mean trajectory length
$\bar\tau=2$ ($32^3$) or $\bar\tau=1$ ($48^3$) was used, and propagator measurements taken
at the end of every trajectory using a randomly chosen point source.
The results presented arise from O(5000) measurements at $j\tilde a=0.005$ up to
O($3\times10^4$) at $j\tilde a=0.05$.
Since the conventional staggered fermion mass term is absent from 
(\ref{eq:Dlatt}), the global symmetry 
\begin{equation}
\psi\mapsto e^{i\alpha\varepsilon(x)}\psi;\;\;
\bar\psi\mapsto e^{i\alpha\varepsilon(x)}\bar\psi;\;\;
\phi\mapsto e^{-i\alpha\varepsilon(x)}\phi;\;\;
\bar\phi\mapsto e^{-i\alpha\varepsilon(x)}\bar\phi
\end{equation}
with $\varepsilon(x)\equiv(-1)^{x_0+x_1+x_2}$, means that $C_N(k,t)$ vanishes
for $t$ even and $C_A(k,t)$ for $t$ odd. Moreover only $\mbox{Re}C_N$ and
$\mbox{Im}C_A$ survive the ensemble average. The resulting correlators are then
fitted to the forms~\cite{Hands:2001aq}
\begin{eqnarray}
C_N(k,t)&=&Ae^{-E_Nt}+Be^{-E_N(L_t-t)};
\label{eq:corrfitsN}\\
C_A(k,t)&=&C(e^{-E_At}-e^{-E_A(L_t-t)}).
\label{eq:corrfitsA}
\end{eqnarray}
to yield the $k$-dependent amplitudes $A$, $B$, $\vert C\vert$ and the energy
$E$, which is extracted independently from both $C_N$ and $C_A$. Stable fits
for $C_N(t)$ and $C_A(t-1)$ were found for the windows $ta_t\in[7,25]$ ($32^3$) and
$ta_t\in[7,41]$ ($48^3$). It is slightly unusual to have to deal with excited state
contamination when fitting a fermion propagator, which may be a symptom of the
strong fluctuations near a QCP; this was found to be an issue particularly for
$k\gapprox k_F$.

\begin{figure}[thb]
\begin{centering}
\includegraphics[width=.7\textwidth]{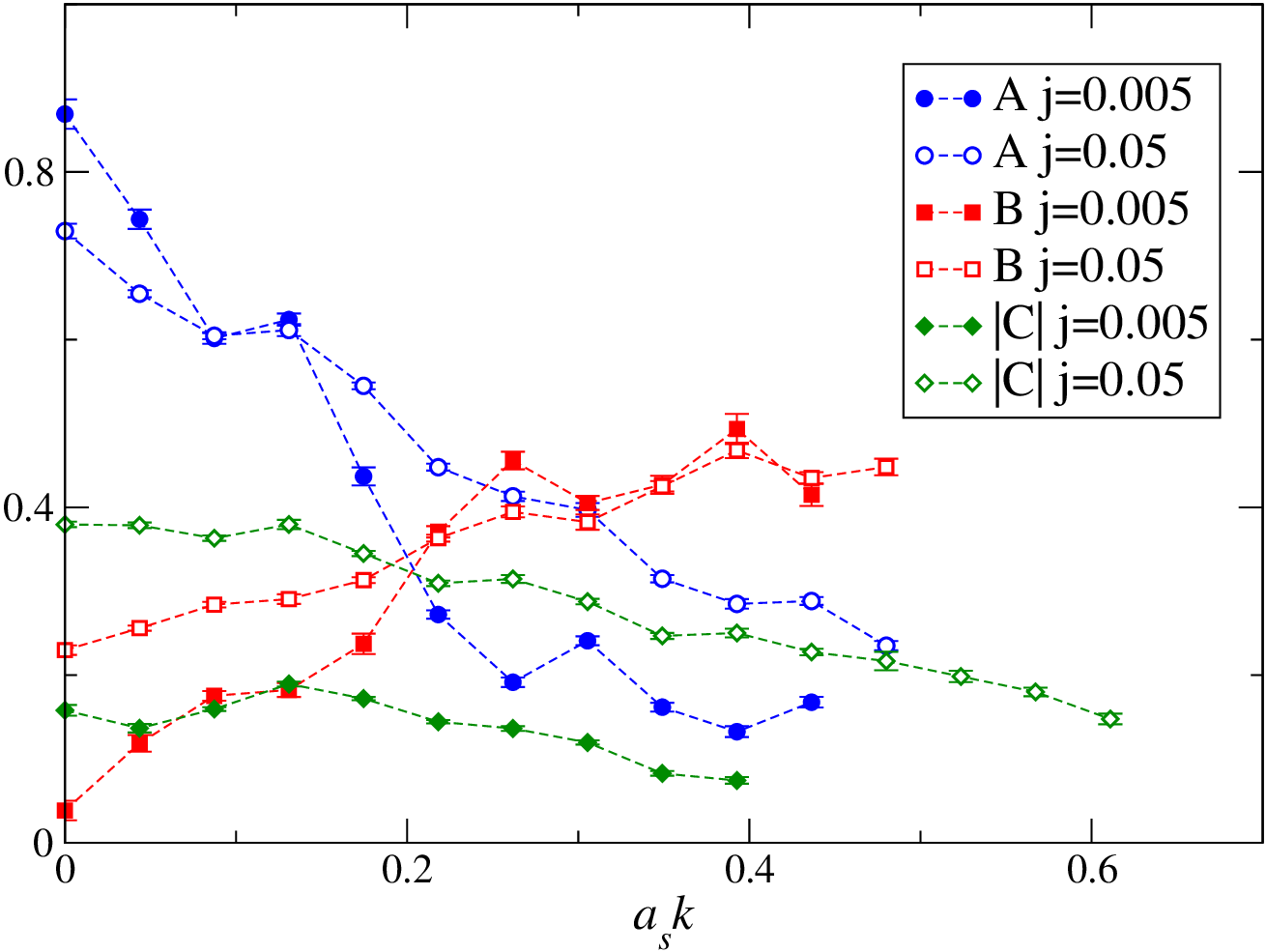}
\caption{Amplitudes obtained from fits to (\ref{eq:corrfitsN},\ref{eq:corrfitsA}) on $48^3$ at
$\mu a_t=0.1$.}
\label{fig:amplitudes}
\end{centering}
\end{figure}
It is instructive first to consider the fitted amplitudes:
Fig.~\ref{fig:amplitudes} shows results from $\mu a_t=0.1$ on $48^3$, for two
values of the exciton
source $j$. In the normal channel the time-asymmetric form of $C_N$ is manifest,
and changes in character as $k$ increases. For small $k$ the forwards
propagating signal is stronger, but $B/A$ grows with $k$ and for $ka_s\gapprox0.2$
the backwards signal dominates. The interpretation is
as follows~\cite{Hands:2001aq}: for $k<k_F$ the dominant excitations are hole-like, and for $k>k_F$
particle-like. Increasing $j\tilde a$ from 0.005 to 0.05 has the effect of smearing the
Fermi surface so that quasiparticles tend to become an admixture of both,
and the disparity between $A$ and $B$ diminishes. 
The effect of smearing is also seen in the anomalous channel, where $C_A$ grows steadily in
magnitude with increasing $j$. In weakly-coupled models~\cite{Hands:2001aq,
Hands:2004uv} $\vert C(k)\vert$ is non-monotonic with a maximum near $k_F$ where
particle-hole mixing is strongest, but
here the behaviour is less clear-cut. The same trends with both $k$ and $j$ are
observed at larger $\mu$.

\begin{figure}[thb]
\begin{centering}
\includegraphics[width=.7\textwidth]{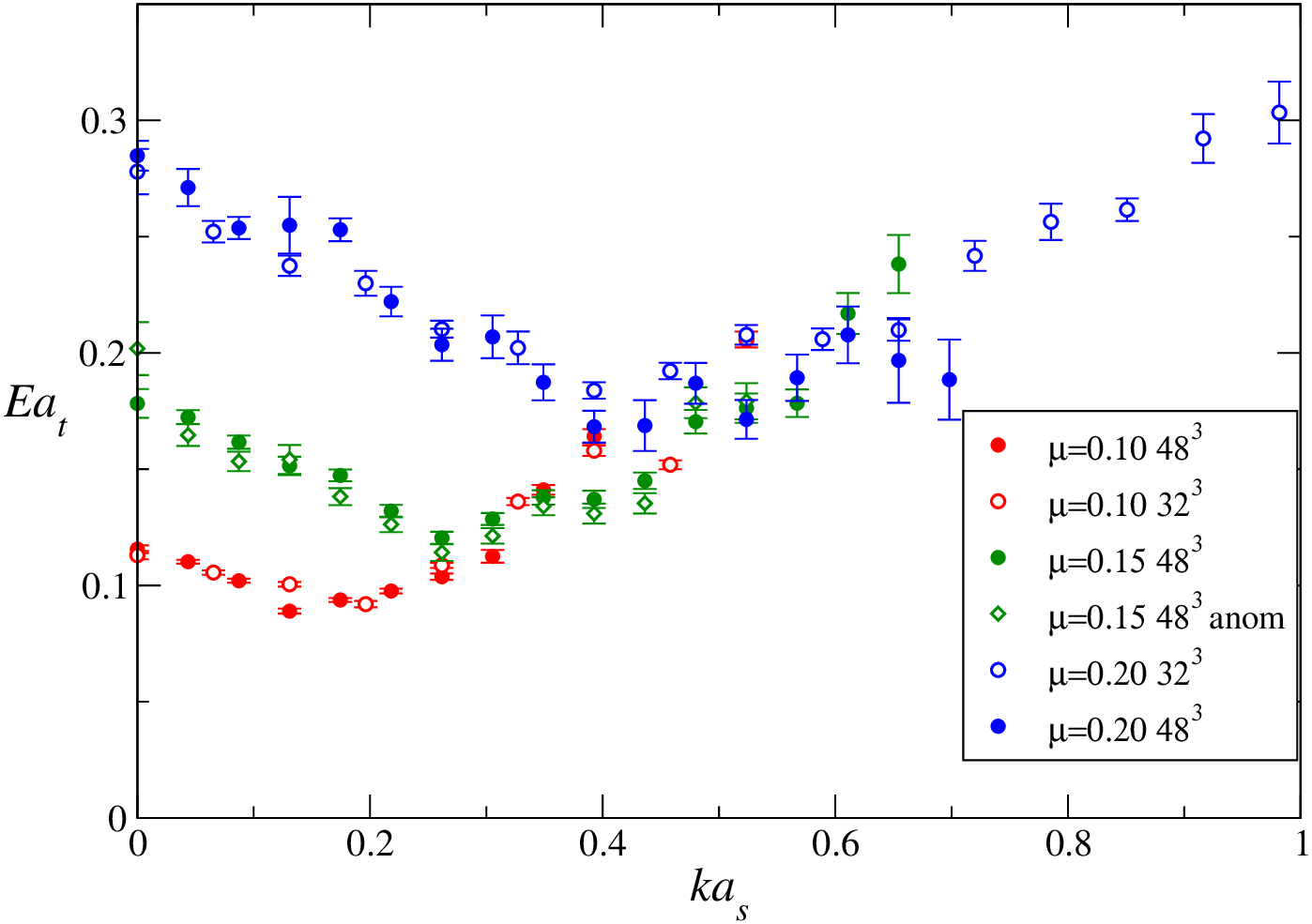}
\caption{Dispersion relation $E(k)$ in the normal channel (unless stated) for
various $\mu$, for fixed $j\tilde a=0.01$.} 
\label{fig:disp_mu}
\end{centering}
\end{figure}
Fig.~\ref{fig:disp_mu} shows results from fits to the dispersion
$E(k)$ for various $\mu$ at $j\tilde a=0.01$ from (\ref{eq:corrfitsN},\ref{eq:corrfitsA}). 
The common feature is that $E(k)$ is
non-monotonic with a minimum in the neighbourhood where the
amplitude ratio
$A/B\approx1$, which we have identified as the Fermi momentum $k_F$. For $k<k_F$
quasiparticle excitations are hole-like, and the energy needed to excite them
from the ground state decreases as $k\nearrow k_F$. For $k>k_F$,
excitations are particle-like and the opposite holds true: indeed, in this
regime results from
all three $\mu$-values studied are plausibly consistent with being drawn from
the same branch of the dispersion curve appropriate to the vacuum (ie. with zero
bias voltage). Fig.~\ref{fig:disp_mu} compares results from two volumes $32^3$
and $48^3$, and also for $\mu a_t=0.15$ with fits in both normal
(\ref{eq:corrfitsN}) and anomlaous (\ref{eq:corrfitsA}) channels. On the assumption that a
smooth curve may be drawn through the admittedly noisy data, there is no
evidence for any significant finite volume artifacts, or systematic difference
between the two channels. 

\begin{figure}[thb]
\begin{centering}
\includegraphics[width=.7\textwidth]{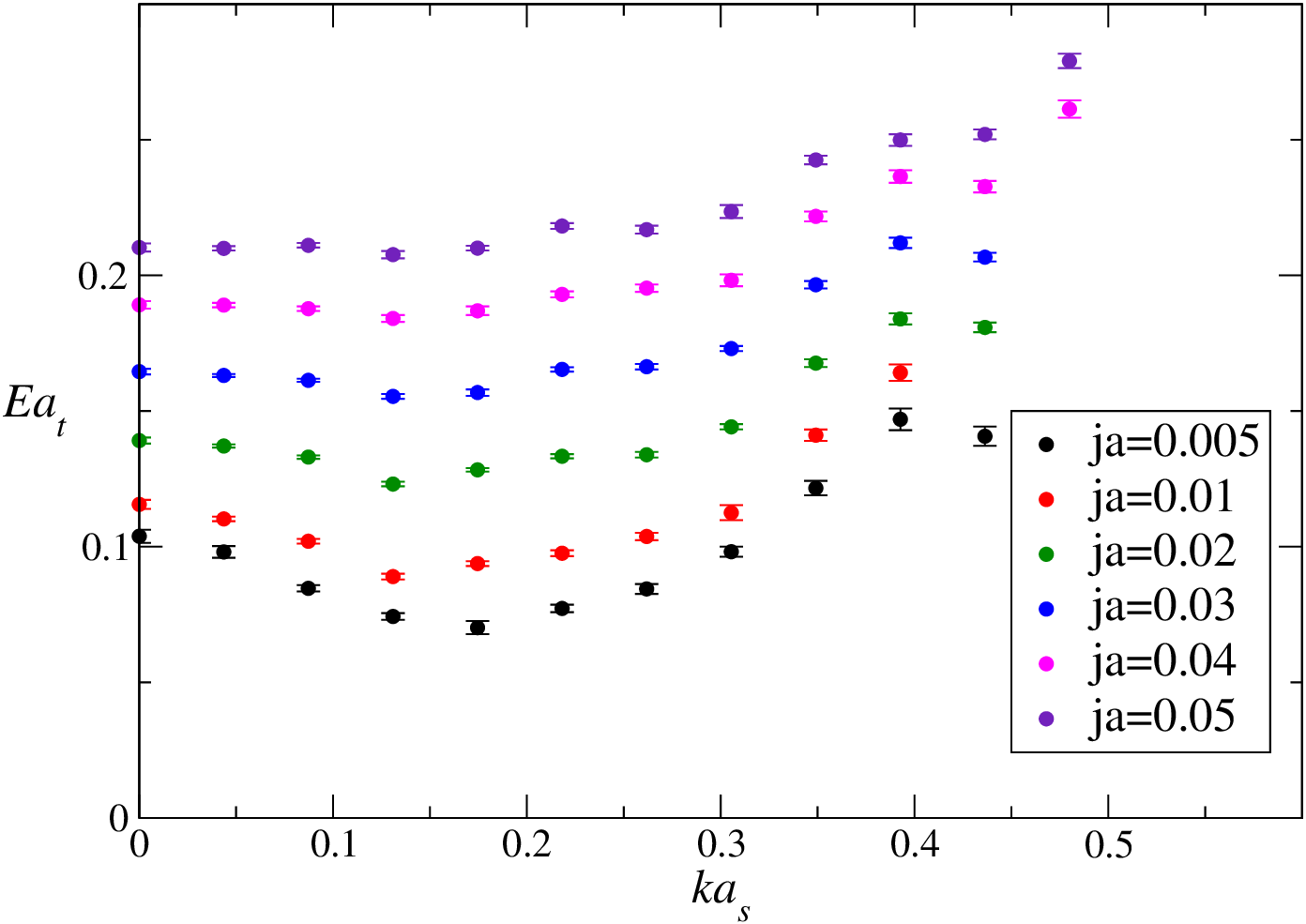}
\caption{$E(k)$ in the normal channel for
$\mu a_t=0.1$, for various $j$ on $48^3$.} 
\label{fig:disp_n_48_mu0.10}
\end{centering}
\end{figure}
Fig.~\ref{fig:disp_n_48_mu0.10} plots $E(k)$ for various $j$ at $\mu a_t=0.1$ on
$48^3$.  The non-monotonicity observed above becomes more pronounced as $j\to0$,
and since there is little shift in the minimum with $j$, we adopt the
pragmatic procedure of identifying the Fermi momentum $k_F$ with the value of
$k$ where $E$ is minimum.  The resulting estimates are shown in
Table~\ref{tab:luttinger}; the quoted error is half the mode spacing on
$48^3$, except at $\mu a_t=0.2$ where the dispersion is flatter and a full mode spacing
is taken. 
\begin{table}[h]
\centering
\setlength{\tabcolsep}{0.4pc}
\hspace{-20mm}
\begin{tabular}{|l|llll|}
\hline
$\mu a_t$ & $k_Fa_s$ & $n_c^{\rm free}(\mu)a_s^2$  & $n_c^{\rm free}(k_F)a_s^2$ & $n_c(\mu,j\to0)a_s^2$ \\
\hline
0.10  & 0.175(22) & 0.0032 & 0.011(3) & 0.0095(1) \\
0.15  & 0.262(22) & 0.0080 & 0.023(4) & 0.0328(2) \\
0.20  & 0.436(44) & 0.0139 & 0.068(15) & 0.0905(3) \\
\hline
\end{tabular}
\caption{Estimates for the Fermi momentum $k_F$ and comparison with Luttinger's
theorem.}
\smallskip
\label{tab:luttinger}
\end{table}
The first observation is that $k_F$ is systematically greater than $\mu$,
consistent with the precocious saturation of $n_c(\mu)$ observed in
\cite{Armour:2013yk}. This is discussed further in Sec.~\ref{sec:discussion},
but already we note a discrepancy with the expectation
$k_F=\mu$ for free massless fermions with $a_s=a_t$. Setting
aside the issue of the smearing of the Fermi surface by an exciton gap $\Delta>0$, we can at this
stage test consistency with Luttinger's theorem, which states that 
$n_c$ depends solely on the geometry of the Fermi surface characterised by
$k_F$, independent of the nature of the interactions. The third column of Table
\ref{tab:luttinger} shows the carrier density $n_c^{\rm free}$ evaluated for free massless fermions
on the same $48^3$ lattice at the reference value of $\mu$, and the fourth 
with $\mu$ set  equal to $k_F$ in the second
column; the fifth column gives the value of $n_c$ in the interacting theory in the limit $j\to0$ obtained
in \cite{Armour:2013yk}. In all cases $n_c^{\rm free}(\mu)\ll n_c^{\rm
free}(k_F)$, while the values of $n_c^{\rm free}(k_F)$ and $n_c(\mu)$ 
are comparable, which is encouraging;  however as $\mu$ increases
$n_c^{\rm free}(k_F)<n_c(\mu)$ indicative of the difficulties in precisely locating $k_F$
due to 
both the limited momentum resolution and perhaps the absence of a
sharp Fermi surface due to exciton condensation.

\begin{figure}[thb]
\begin{centering}
\includegraphics[width=.7\textwidth]{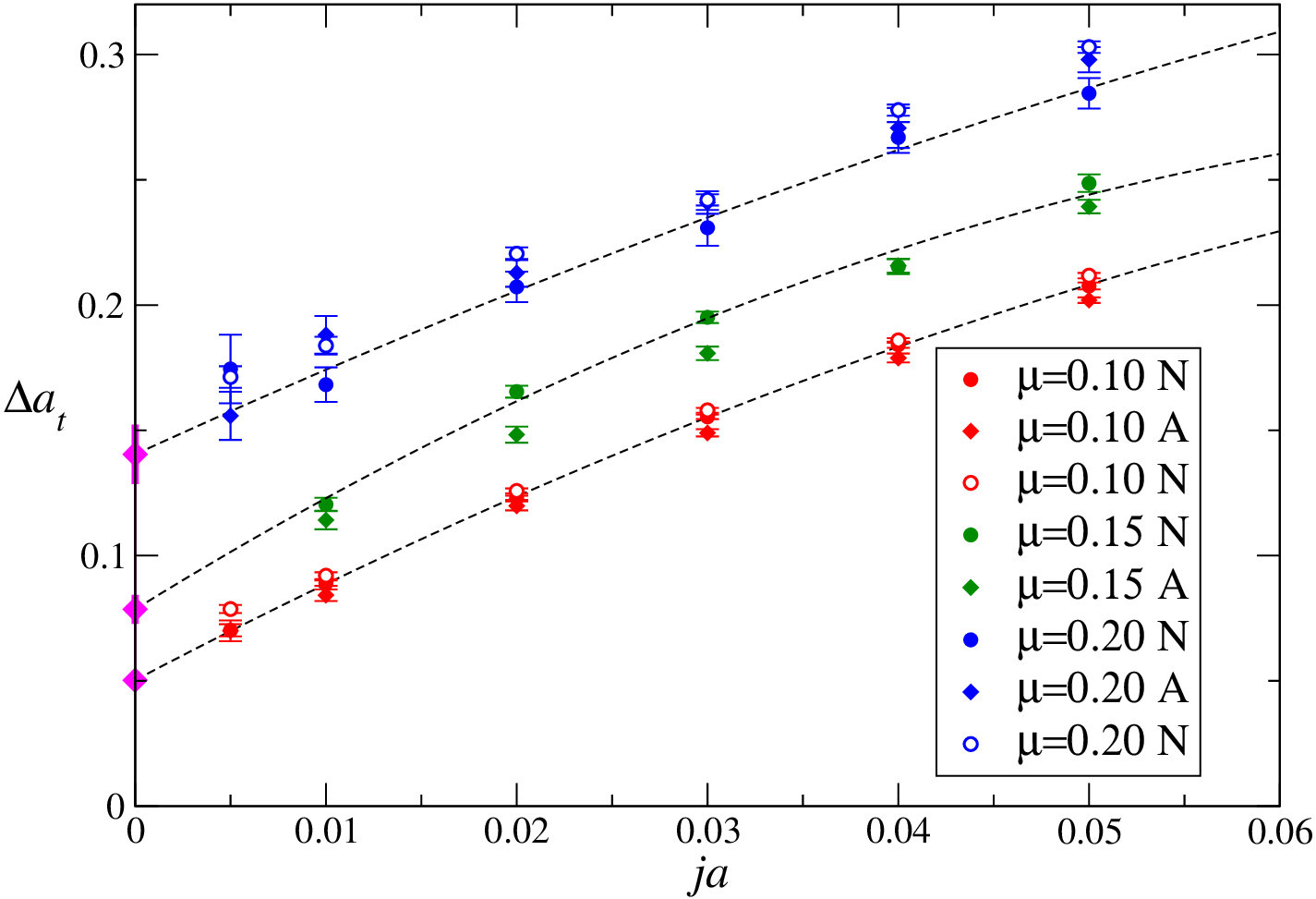}
\caption{Gap $\Delta$ vs. $j$ for various $\mu$. Filled symbols are from $48^3$,
empty from $32^3$, and the dotted lines are fits
to normal data on $48^3$ using (\ref{eq:quadfit}).}
\label{fig:gap_j}
\end{centering}
\end{figure}
In Fig.~\ref{fig:gap_j} we plot the gap $\Delta$ defined as the energy
$E(k_F)$. Data from all available fits on both volumes is shown, in
both normal and anomalous channels. As before there is little evidence for a
systematic effect with lattice volume and both channels yield consistent results.
The dashed lines show an extrapolation to $j=0$ based on the quadratic form
\begin{equation}
\Delta=\Delta_0+aj+bj^2.
\label{eq:quadfit}
\end{equation}  
It is clear $\lim_{j\to0}\Delta(j)\not=0$ which is direct evidence for
spontaneous gap formation via exciton condensation for all three values of
$\mu$.
\begin{figure}[thb]
\begin{centering}
\includegraphics[width=.7\textwidth]{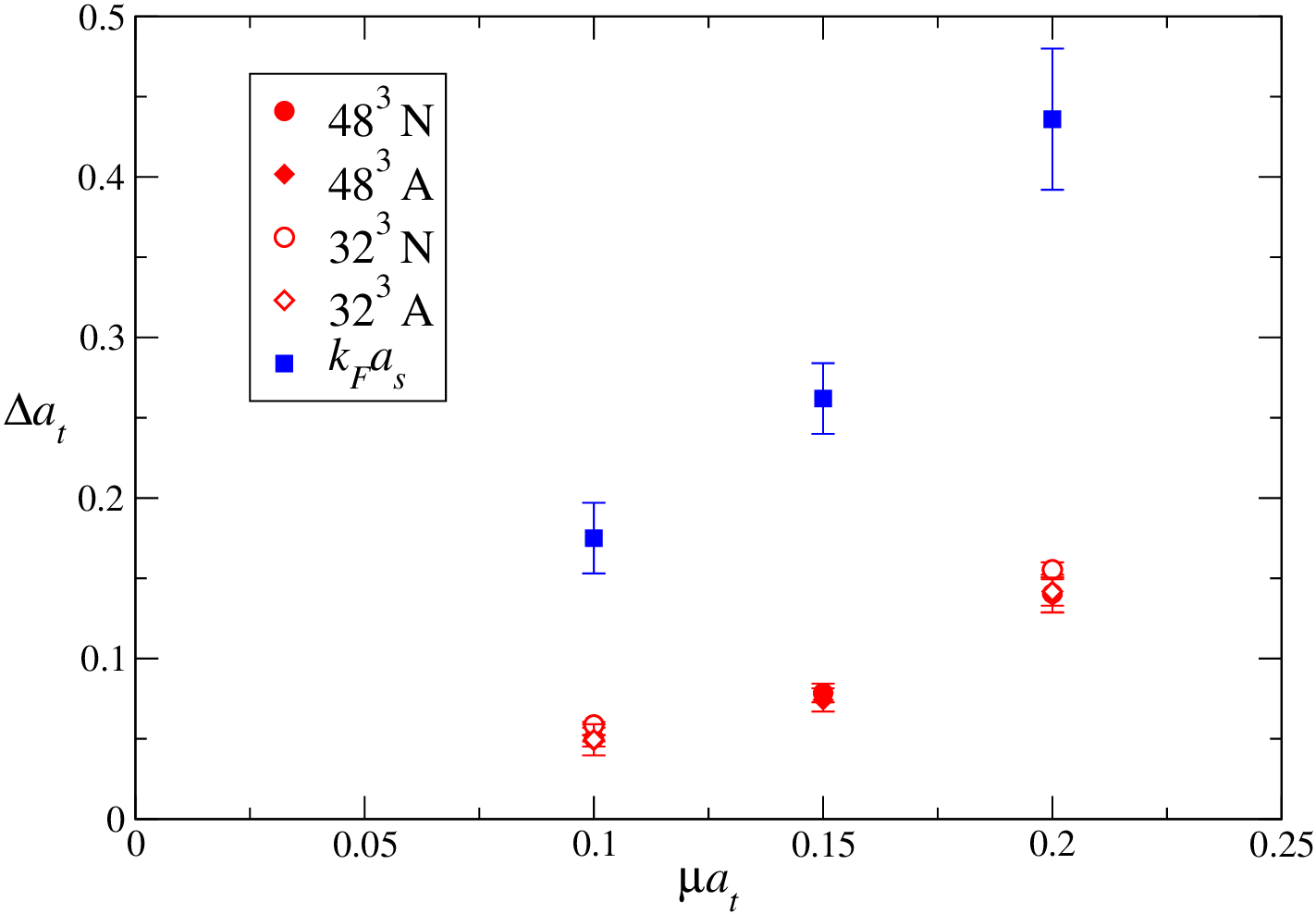}
\caption{$\Delta(j=0)$ versus $\mu$. The square symbols denote $k_F$ from
Table~\ref{tab:luttinger}.} 
\label{fig:gap}
\end{centering}
\end{figure}
The resulting extrapolations, together with the estimates for $k_F(\mu)$ from
Table~\ref{tab:luttinger} are shown in Fig.~\ref{fig:gap}, which is the main
result of this paper. In the region of $\mu$ studied 
it is clear that $\Delta$ varies strongly with and is of
the same order of magnitude as the chemical potential $\mu$, which are notable 
features of this particular model proposed in \cite{Armour:2013yk}, and indicative
of strong interactions at the Fermi surface. This conclusion holds in both
normal and anomalous channels, appears to be independent of volume, and is striking enough
to be 
robust against uncertainties introduced by the {\it ad hoc\/} nature of our
analysis (an analytic form against which to fit the dispersion $E(k)$ would be
very valuable), the IR and UV artifacts inherent in studies of lattice
models with $\mu\not=0$~\cite{Cotter:2012mb}, and lack of knowledge of the physical anisotropy
$a_t/a_s$.  On the basis of the three chemical potentials studied 
both $\Delta$ and $k_F$ appear to scale superlinearly with $\mu$, in qualitative
agreement with (\ref{eq:Deltahyp}).

\section{Discussion}
\label{sec:discussion}

In this paper we have used lattice Monte Carlo simulation techniques to explore
the quasiparticle dispersion relation in an interacting field theory with
non-zero charge density, and shown that for $k\sim k_F$ the excitations are
gapped with $\Delta\sim O(\mu)$, and $\Delta$ scaling faster than linearly with
$\mu$. This is in sharp contrast to results from
comparable studies of other simulable models with $\mu\not=0$. In
\cite{Hands:2004uv} the gap in the 3+1$d$ Nambu Jona-Lasinio model (a
relativistic analogue of the original BCS model) was shown to be approximately
constant above onset, independent of and numerically much smaller than $\mu$, consistent with the BCS
result  $\Delta\sim\Lambda_{UV}\exp(-c\Lambda_{UV}^2/\mu^2)$. 
In QCD with gauge group SU(2) there is a
so-called {\it quarkyonic\/} regime above onset where
$\langle\Psi\Psi\rangle\propto\mu^2$~\cite{Cotter:2012mb}; this is consistent
with degenerate fermions in 3+1$d$ with a gap $\Delta\sim O(\Lambda_{QCD})$ 
independent of $\mu$. It is also very different from the result
$\Delta/\mu\sim O(10^{-7})$ obtained by self-consistent diagrammatic
techniques~\cite{Kharitonov}, although comparable with the large values of
$\Delta/\mu$ obtained in \cite{SPM}, where it was found that 
$\Delta$ depends sensitively on the treatment of screening effects, and in
particular on the reduction of screening once the superfluid gap forms.
One feature of our approach which does merit comparison with the treatment in
\cite{SPM} is that competition between inter- and
intra-layer pairing condensates can be addressed; see Fig.~11 of \cite{Armour:2013yk}.

We now return to Table~\ref{tab:luttinger} and the issue of why
$\mu a_t<k_Fa_s$. In \cite{Armour:2013yk} it was suggested this is because in a
strongly self-bound system the Fermi energy is necessarily less than the Fermi momentum
in natural units. While this may be plausible for a system
where $\mu\gg\Delta$ is by far the largest scale,
it is difficult to see how this picture can persist in 
the regime we have been focussing on. Another possibility, which we cannot
dismiss, is that the disparity is a lattice artifact caused by a
large induced anisotropy $a_t/a_s\sim O(0.5)$. Indeed, if we assume the Fermi
velocity remains close to one even in the presence of interactions, then the
dispersion data of Fig.~\ref{fig:disp_mu} might suggest $a_t/a_s\approx0.3$.
However, a more compelling possibility is that $E(k)$ is not a linear
relation, but rather a power law characteristic of a nearby QCP. 
Rewrite the scaling form (\ref{eq:nscaling})
as $n_c\propto\mu^{2\over z}$; the Luttinger scaling $n_c\propto k_F^2$ then
gives $E\propto k^z$, where guided by Fig.~\ref{fig:disp_mu} we assume the
relation between $E_F$ and $k_F$ completely characterises the quasiparticle
dispersion. In this scenario the scaling (\ref{eq:nscaling}) extracted from bulk observables in
\cite{Armour:2013yk} thus yields an estimate for the dynamical critical exponent 
\begin{equation}
z\approx0.6.
\end{equation}

Much greater numerical precision than achieved in
Fig.~\ref{fig:disp_mu} would be
needed to distinguish these two possibilites unambiguously. Another route would
be to perform a ``biased bilayer'' study for a related 2+1$d$ theory, the
Thirring model~\cite{DelDebbio:1997dv}, 
whose behaviour as a function of $N_f$ and $\mu$ is 
qualitatively similar to the model here~\cite{Christofi:2007ye}, but whose continuum action is
manifestly covariant implying $a_t\equiv a_s$ throughout.

Finally, we note that the superlinear scaling of $\Delta(\mu)$ is also
suggestive of a power law $\Delta\propto\mu^\sigma$, with $\sigma>1$, 
modifying our naive expectation $\Delta\propto\mu$. Clearly, a much more
extensive simulation campaign is required to verify this interesting
possibility.

\section{Acknowledgements}

The authors would like to acknowledge the use of computing resources at Diamond
Light Source and the University of Oxford
Advanced Research Computing (ARC) facility 
({\tt http://dx.doi.org/10.5281/zenodo.22558})
in carrying out this work.
We estimate
approximately 1 million core hours of computing time were needed. 
CPUs used were either Intel Xeon E5-2640v3 Haswell or E5-2650 SandyBridge. 
SJH was supported by STFC grant ST/L000369/1, and thanks Allan MacDonald for
helpful discussions.

\end{document}